\documentclass[12pt]{iopart}
\usepackage{graphicx}
\usepackage{subfigure}
\usepackage{color}
\usepackage{ulem}

\def\al{\alpha}
\def\be{\beta}
\def\ga{\gamma}
\def\de{\delta}

\def\ga{\gamma}

\def\la{\lambda}
\def\si{\sigma}
\def\th{\theta}
\def\mn{{\mu\nu}}
\def\Ga{\Gamma}

\def\prt{\partial}
\def\lrdrt#1{\hskip-2pt\stackrel{\leftrightarrow}{D_{#1}}\hskip-2pt}
\def\scroodle#1{\vbox{\ialign{##\crcr\notsimfill\crcr
  \noalign{\kern-4pt\nointerlineskip}
   $\hfil\displaystyle{#1}\hfil$\crcr}}}

\def\twiddle{\lower4pt\hbox{\hskip-0pt{$\widetilde{}$}}}
\def\m@th{\mathsurround=0pt}
\def\cmapstochar{\mathrel{\rlap{
  \lower0.1pt\hbox{\hskip-1.75pt{$\mapstochar$}}}
  \raise0pt\hbox{\hskip2.5pt{$\twiddle$}}}}
\def\notsimfill{$\m@th\cmapstochar$}

\def\Xtw{\scroodle{t}_{\la \mn \ldots}}

\newcommand{\bea}{\begin{eqnarray}}
\newcommand{\eea}{\end{eqnarray}}
\newcommand{\bit}{\begin{itemize}}
\newcommand{\eit}{\end{itemize}}
\newcommand{\hf}{\frac{1}{2}}
\newcommand{\ie}{{\it i.e.}}

\newcommand{\eg}{{\it e.g.}}
\newcommand{\nn}{\nonumber\\}

\providecommand{\Journal}[4] {#1 {\bf#2}, #4 (#3)}
\providecommand{\APB}{Appl. Phys. B} %
\providecommand{\ARNPS}{Annu. Rev. Nucl. Part. Sci.} %
\providecommand{\CRASP}{C.R. Acad. Sci. (Paris)}%
\providecommand{\CQG}{Class. Quantum Grav}%
\providecommand{\EPJC}{Eur. Phys. J. C.} %
\providecommand{\HI}{Hyperfine Interact.} %
\providecommand{\JCAP}{J. Cosm. Astro. Phys.}%
\providecommand{\JMP}{J. Math. Phys.} %
\providecommand{\JPCS}{J. Phys. Conf. Ser.}%
\providecommand{\LNP}{Lect. Notes Phys.} %
\providecommand{\NAT}{Nature} %
\providecommand{\NaP}{Nature Phys.} %
\providecommand{\PPNL}{Phys. Part. Nucl. Lett.}
\providecommand{\PR}{Phys. Rev.} %
\providecommand{\PRL}{Phys. Rev. Lett.} %
\providecommand{\PRA}{Phys. Rev. A} %
\providecommand{\PRD}{Phys. Rev. D} %
\providecommand{\RMP}{Rev. Mod. Phys.} %
\providecommand{\PLB}{Phys. Lett. B} %
\providecommand{\PT}{Phys. Today} %
\providecommand{\SCI}{Science}%
\providecommand{\IJMPCS}{Int. J. Mod. Phys. Conf. Ser.} %
\begin{document}

\title[The CPTV effects on neutron gravitational bound state]
{The CPT-violating effects on neutrons' gravitational bound state}

\author{Zhi Xiao$^1$\footnote{Corresponding author.} and Lijing Shao$^2$}

\address{$^1$ Department of Mathematics and Physics, North China Electric Power University, Beijing 102206, China}\ead{spacecraft@pku.edu.cn}

\address{$^2$ Kavli Institute for Astronomy and Astrophysics, Peking University, Beijing 100871, China}

\vspace{10pt}
\begin{indented}
\item[]January 2020
\end{indented}

\begin{abstract}
In this work, the CPT-violating (CPTV) interactions on neutrons' gravitational bound state
are studied. With simple analytical solutions, we provide a preliminary investigation on the Lorentz-violation (LV) induced spin precession
due to the $\vec{\sigma}\cdot\vec{\tilde{b}}(1+gz)$ and $\bar{b}/m_{_I}\vec{\sigma}\cdot\hat{\vec{p}}$ couplings, where $\vec{\tilde{b}}$ and $\bar{b}$ represent LV coefficients.
The helicity-dependent couplings can induce unusual phase evolutions with position and momentum dependence.
As $\vec{\tilde{b}}$ varies with time due to the Earth's motion, the spin polarization also shows a
sidereal time dependence, and it may be enhanced with time for ultra-stable polarized state of neutrons.
The inseparability of the spin-momentum coupling of the $\bar{b}$-term can also lead to motional dependent
polarization state.
With the precisely measured transition frequency between different gravitational bound states, we get
a rough bound $|\vec{\tilde{b}}|<3.9\times10^{-3}$GeV for unpolarized neutrons.
If the spin-flip transition frequency can reach comparable precision in the future, the bound can
be improved to the level of $10^{-24}$GeV.
The test of weak equivalence principle with polarized atom may also improve it significantly.
\end{abstract}


\submitto{\jpg}
\maketitle

\section{Introduction}
Symmetry and its breaking pattern are the main theme of physics in the last century. On one hand, the $\mathrm{SU(3)}_C\otimes\mathrm{SU(2)}_L\otimes\mathrm{U(1)}_Y$ symmetry is responsible for the observed electroweak and strong forces dominant in the microscopic world, while local Lorentz and diffeomorphism invariance are responsible for the gravity
dominant in the macroscopic world. All the fundamental forces are closely embedded into the gauge structure. On the other hand, scalar bosons arising from the spontaneous breaking of certain (approximate) global symmetries can also account for forces we have in nature, such as the nuclear force due to the exchange of $\pi$-meson. However, many puzzles still remain challenge to our current
paradigm of understanding, such as dark matter and dark energy,
not to mention the still on-going searching for a satisfactory theory of quantum gravity.
A very interesting scenario proposed in Ref.~\cite{AKRB} is that the Lorentz symmetry may be spontaneously broken, and
this may help to resolve the long standing puzzles \cite{LVDE} mentioned above.
Moveover, within the framework of the standard model extension (SME), established by Kosteleck\'y and collaborators \cite{SME}\cite{SMEG},
various phenomenological effects arising from tiny Lorentz violation can be studied systematically \cite{DATA}.
This has spurred extensive investigations on the test of Lorentz symmetry in the last two decades \cite{AKCPT}.

The continuous investigations of the SME in flat space have already deepened into the non-minimal (power-counting nonrenormalizable) sector \cite{Nonmini}, and extensive exploration on interacting (not only kinematic) Lorentz-violating (LV) operators has also been carried out \cite{InteractLV}.
In the curved spacetime, Refs.~\cite{Quentine} and \cite{CounterShade}\cite{MGC} provide systematic and elegant theoretical analyses
on LV couplings in the pure gravity and matter sectors, respectively.
The non-relativistic (NR) spin-independent fermion-gravity couplings in general gravitational field have been extensively studied in Ref.~\cite{MGC}, and the NR spin-dependent counterparts in the uniform gravitational field have been studied in Ref.~\cite{YuriEPI}.
Due to the neutrality and very tiny polarizability under external electromagnetic fields \cite{Snow-N},
the neutron has long been an excellent candidate to the test of interplay between gravity and quantum mechanics \cite{COW}.
Recently, the spin-independent LV neutron-gravity couplings \cite{Escobar1} and pure LV gravity couplings \cite{Escobar2}
have been thoroughly studied in attempts to analyze the GRANIT experiment, performed in the Institut Laue Langevin (ILL) \cite{UCNILL}.
Very recently, Abele \etal.~have also systematically studied the effects of LV neutron couplings \cite{IWAN} with the conventional uniform gravitational field, $mgz$, in the NR context \cite{AKCL}.
To our best knowledge, an extensive study of spin-dependent LV fermion-gravity couplings is still under development \cite{AKZL}.
In this work, instead of treating the uniform gravitational field as a pure external field \cite{IWAN},
we try to provide simple analytical case studies on the CPT-violating spin-dependent neutron couplings interwoven with the linear gravitational potential.
This may provide a simple investigation on the much wider variety of spin-dependent LV effects caused by the LV matter-gravity couplings \cite{MGC}\cite{AKZL}.
As is well-known, spin-gravity effects for subatomic particles are very tiny even in the Lorentz invariant (LI) context \cite{SpinRota1}\cite{WTNH}.
However, the fast development in precision measurement is quite impressing \cite{AO}\cite{SBDKDC},
and another source of gravity may couple is, besides the energy-momentum, spin \cite{Torsion},
which makes the search for spin-gravity effects \cite{Obukohov} indispensable.

Moreover, a subset of operators in the SME share essentially the same forms as those originating in other exotic scenarios,
such as torsion and nonmetricity \cite{Torsion}\cite{TNLV}\cite{AKNM}\cite{Snow-SN}.
As an example, the spin-dependent LV $b^\mu$ term can also entail the structure of minimal torsion $T^\la_{~\mn}$ coupling
if we identify $b^\mu_\mathrm{eff}\equiv b^\mu-\frac{1}{8}\epsilon^{\mu\al\be\ga}T_{\al\be\ga}$ \cite{MGC}.
With stringent bounds on torsion and nonmetricity available \cite{TNLV}\cite{AKNM}\cite{Snow-SN},
we simply assume our analysis below is based on a torsion and nonmetricity free context.

The organization of this work is the following.
By focusing on the spin-dependent operators in Ref.~\cite{YuriEPI}, we obtain an effective Hamiltonian for vertical motion
in Sec.~\ref{Preparation} using the reduction ansatz in Ref.~\cite{Escobar1}. As a case study on the LV spin-gravity effects, tiny LV correction to the Larmor frequency from the spin evolution equation is obtained.
In Sec.~\ref{QMLinear}, we review the eigen-solution in the linear gravitational potential.
In Sec.~\ref{QMLVb}, with {\it ad hoc} assumptions, we provide analytical solutions of gravitational bound states and discuss the spin precession and polarization evolution in the presence of $\vec{\tilde{b}}$-type couplings.
At last, we utilize the quantum perturbation theory to give the leading order frequency shifts due to the CPTV spin-gravity coupling,
which coincides with the approximation of the exact eigen-energy obtained in Eq.~(\ref{ExactEn}).
In Sec.~\ref{PersDis}, we summarize our results and speculate on possible experiments
which may be potentially viable to the test of the CPTV spin-gravity couplings,
such as the weak equivalence principle test with polarized matter.

\section{The Hamiltonian with LV fermion-gravity couplings}\label{Preparation}
\subsection{Hamiltonian in the uniform gravitational field}

The general action \cite{SME}\cite{SMEG} describing the LV neutron-gravity coupling is
\bea\label{matterAct}&&
\hspace{-10mm}S_\psi=\int{d^4x}e \left[\frac{i}{2}e^\mu_{~a}\bar{\psi}\Ga^a{\lrdrt\mu}\psi-\bar{\psi}M\psi \right],\\\label{GGamma}&&
\hspace{-10mm}\Ga^a\equiv\ga^a-\left[c_{\mn}\ga^b+d_{\mn}\ga_5\ga^b\right]e^{\nu a}e^\mu_{~b}-\left[e_\mu+if_\mu\ga_5\right]{e^{\mu a}}
-\hf g_{\la\mn}e^{\nu a}e^\la_{~b}e^\mu_{~c}\si^{bc},\\\label{GMass}&&
\hspace{-10mm}M\equiv{m+a_\mu e^\mu_{~a}\ga^a+b_\mu e^\mu_{~a}\ga_5\ga^a+\hf H_{\mn}e^\mu_{~a}e^\nu_{~b}\si^{ab}},
\eea
where $\bar{\chi}\Ga^a{\lrdrt\mu}\psi\equiv\bar{\chi}\Ga^aD_\mu\psi-(\bar{\chi}\bar{D_\mu})\Ga^a\psi$, $D_\mu\psi\equiv\left[\prt_\mu+\frac{i}{4}\omega_\mu^{ab}\si_{ab}\right]\psi$
and $\bar{\chi}\bar{D_\mu}\equiv\prt_\mu\bar{\chi}-\frac{i}{4}\omega_\mu^{ab}\bar{\chi}\si_{ab}$.
The first terms in (\ref{GGamma}) and (\ref{GMass}) are the usual gamma matrix $\ga^a$ and Lorentz invariant mass $m$ respectively,
while all the other terms with LV coefficients fields $a_\mu,~b_\mu,~c_{\mn},~d_{\mn},~e_{\mu},~f_{\mu},~g_{\la\mn},~H_{\mn}$
lead to LV matter-gravity couplings.
For consistency, these LV coefficients fields have to be position dependent \cite{SMEG} in the presence of gravity.
In other words, for a generic LV coefficients field $t_{\la\mn...}=\bar{t}_{\la\mn...}+\Xtw$, where $\bar{t}_{\la\mn...}$
is the vacuum expectation value, and $\Xtw$ is the fluctuation. However, in the test particle assumption,
the contribution of these fermion fields to the energy-momentum can be ignored, and this avoids the consistent problem in Einstein field equation. As a consequence, the fluctuations of LV coefficients fields can also be ignored in this context.
Another way to view the constant LV coefficients is that for an approximation of the linear gravitational potential, which
will be adopted in the following discussions, it is equivalent to the linear acceleration and the spacetime is essentially flat.
In the following, what we mean LV coefficients will exclusively refer to their vacuum expectation values, and therefore are constants.
For simplicity, we instead use unbarred notation of LV coefficients in the following.
The $e^\mu_{~a}$ and $\omega_\mu^{ab}$ are the vierbein and spin connection, and $e$ is the
determinant of vierbein.
The convention is the same as in Ref. \cite{SMEG}, with the signature $\mathrm{diag} \, \eta_{\mn}=(-1,1,1,1)$ and $\epsilon_{0123}=1$.
Assuming torsion free, $T^\la_{~\mn}=0$, and choosing the
Schwinger gauge, $e^0_\rho=(1+\vec{g}\cdot\vec{x})\de^0_\rho$ and $e^j_\rho=\de^i_\rho$,
one can get the Hamiltonian (13) in Ref.~\cite{YuriEPI} with the test particle assumption.
Though the metric $g_{\rho\si}=\eta_{ij}e^i_{~\rho} e^j_{~\si}$ is essentially flat, the linear potential
$\vec{g}\cdot\vec{x}$ describes the gravitational field near the Earth surface to a good approximation.

By performing the Foldy-Wouthuysen transformation \cite{MGC}\cite{YuriEPI}\cite{FW}\cite{IGMQM} order by order,
the non-relativistic spin-dependent Hamiltonian, Eq.~(38) in Ref.~\cite{YuriEPI}, can be obtained.
The LI part of the Hamiltonian \cite{WTNH} is
\bea\label{SpinLI}&&
\hspace{-10mm}\hat{H}_\mathrm{LI}=m(1+\Phi)+\frac{(1+\Phi)\hat{\vec{p}}^2}{2m}-\frac{i}{2m}\vec{\nabla}\Phi\cdot\hat{\vec{p}}
+\frac{(1+\Phi)}{4m}\vec{\si}\cdot(\vec{\nabla}\Phi\times\hat{\vec{p}}).
\eea
Note $\Phi\equiv{gz}$ can be regarded as a leading-order approximation of the Newtonian potential $-GM/\sqrt{(R_{\oplus}+z)^2+\vec{r}_\perp^2}\simeq-GM/R_{\oplus}(1-z/R_{\oplus})$ up to a constant,
where $z$ is the vertical coordinate, $\vec{r}_\perp\equiv x\hat{e}_1+y\hat{e}_2$ is expressed in the horizontal coordinates
in the laboratory frame, and $R_{\oplus}$ is the Earth radius.
The first three terms in Eq.~(\ref{SpinLI}) correspond to the redshift of the LI energy, and the last term
is the inertial spin-orbit coupling term.

As for the spin-dependent LV couplings, for notational simplicity, we define
\bea\label{bLVeff1}&&
\acute{b}_{k}^{~i}\equiv\de^i_k\hat{b}_0+m\left[\frac{\epsilon_k^{~lm}}{2}g_{lm}^{~~~i}+\epsilon_k^{~il}\hat{g}_{0l0}\right],\quad
\check{b}_k\equiv b_k+\frac{m\epsilon_k^{~lm}}{2}\hat{g}_{lm0},\\
\label{dLVeff}&&
\tilde{d}_{k0}\equiv\hat{d}_{k0}+\frac{\epsilon_k^{~lm}}{2m}H_{lm},\quad \tilde{d}_k^{~i}\equiv d_k^{~i}+\frac{\epsilon_k^{~il}}{m}\hat{H}_{0l}
\eea
following the spirit of Eq.~(27) in Ref.~\cite{FermiNonmini}.
Since we are only interested in the spin-dependent part, the corresponding LV operators \cite{YuriEPI} are
{\small
\bea\label{SpinLV}&&
\hspace{-23mm}\hat{H}_\mathrm{LV}=-\tilde{b}_k\si^k(1+\Phi)+\left[m(\de^i_k\hat{d}_{00}+\tilde{d}_k^{~i})
-\acute{b}_{k}^{~i}\right]\si^k\left[(1+\Phi)\frac{\hat{p}_i}{m}-\frac{i}{2}\frac{\nabla_i\Phi}{m}\right]
+\left[\check{b}_k\eta^{ij}-\de^j_k\b{b}^i\right]\cdot
\nn&&\hspace{-20mm}~~~~\frac{\si^k}{2m^2}\left[(1+3\Phi)\hat{p}_i\hat{p}_j-3i\nabla_{(i}\Phi\hat{p}_{j)}\right]
+\left[2\de^j_k\tilde{d}_{(0i)}+\epsilon_k^{~il}(\hat{g}_{l~0}^{~j}-\hat{g}_{0l}^{~~j})\right]\frac{\si^k}{m}p_{(i}(1+\Phi)p_{j)},
\eea
}
where the terms with a pair of indices in parentheses indicate symmetrization, such as $\tilde{d}_{(0i)}=(\hat{d}_{0i}+\hat{d}_{0i})/2$.
To be consistent with the standard notation in Refs.~\cite{AKCL}\cite{DATA}, we have defined
\bea
\hspace{10mm}\tilde{b}_k\equiv[\check{b}_k-m\tilde{d}_{k0}],\quad
\b{b}_k\equiv[\check{b}_k+m\tilde{d}_{k0}].
\eea
Note the hat on LV coefficient with $n$ $0$-index
means a multiplication by $1-n\Phi$, such as $\hat{d}_{00}\equiv{d}_{00}(1-2\Phi)$; for details, see Ref.~\cite{YuriEPI}.
As the LV coefficients in Eq.~(\ref{SpinLV}), \eg, $\tilde{b}_k$, are corrected by $\Phi=\vec{g}\cdot\vec{x}$, unlike the Minkowski case \cite{AKCL}, they represent LV spin-gravity couplings in the leading approximation.
If we mildly assume that the dimensional LV coefficients, such as $\tilde{b}_k,~m\hat{g}_{l~0}^{~j}$, are roughly of the same order,
the leading- and second-order LV operators are in the first row of Eq.~(\ref{SpinLV}).
In comparison, the operators in the second row are suppressed by $\mathcal{O}[(\frac{p}{m})^2]$, and will be disregarded
in the following.

For clarity, we rewrite the first line in Eq.~(\ref{SpinLV}) as
{\small
\bea\label{HeffBLV}&&
\hspace{-21mm}\hat{H}_\mathrm{b}=-\vec{\si}\cdot\vec{\tilde{b}}(1+\Phi)-\frac{\bar{b}}{m}
\left[(1+\Phi)(\vec{\si}\cdot\hat{\vec{p}})
-\frac{i}{2}(\vec{\si}\cdot\vec{\nabla}\Phi)\right]
-\frac{[b_\mathrm{N}]_{k}^{~i}}{m}\si^k\left[(1+\Phi)\hat{p}_i-\frac{i}{2}\nabla_i\Phi\right],
\eea
}
where we have intentionally separated
{\small
\bea\label{redefbd}&&
\left[\acute{b}_{k}^{~i}-m(\de^i_k\hat{d}_{00}+\tilde{d}_k^{~i})\right]=[b_\mathrm{N}]_{k}^{~i}
+\bar{b}\de^i_k,\nn&&
\delta_i^k[b_\mathrm{N}]_{k}^{~i}=0,\quad ~~
\bar{b}\equiv\hat{b}_0+\frac{m\epsilon_k^{~lm}}{2}g_{lm}^{~~~k}-m\left[\hat{d}_{00}+\frac{\tilde{d}_k^{~k}}{3}\right].
\eea
}
Note $\bar{b}$ contains the pure time-component of $b_\mu$ and $d_{\mn}$ coefficients.
In a broader perspective, the first two terms in Eq.~(\ref{HeffBLV}) share a similar structure
as terms from axion or axionlike particles.
For example, if we identify $\bar{b}\Phi\rightarrow-\frac{f_{12+13}}{4\pi r}$ and the force range $\la\rightarrow+\infty$,
the $\Phi$-dependent part of $\bar{b}$ operator can be identified as the type $12$-th and $13$-th operators in
Refs.~\cite{SBDKDC}\cite{SpinDM}, \ie,
\bea&&
\hspace{-15mm}\frac{f_{12+13}}{8\pi m}\vec{\si}\cdot \left\{\hat{\vec{p}},~\frac{e^{-r/\lambda}}{r}\right\}
=\frac{f_{12+13}}{4\pi m}\frac{e^{-r/\lambda}}{r}(\vec{\si}\cdot\hat{\vec{p}})
+\frac{if_{12+13}}{8\pi m}\frac{e^{-r/\lambda}}{r^2}\left(1+\frac{r}{\lambda}\right)\vec{\si}\cdot\hat{r}.
\eea
Note that the Hamiltonian (\ref{HeffBLV}) can lead to LV corrections to neutrons' spin precession in a weak magnetic field $\vec{B}$.
From the Heisenberg equation,
\bea\label{HeisenEqn}&&
\hspace{-18mm}\frac{d\vec{S}}{dt}=i\left[-\vec{\mu}\cdot\vec{B}+\hat{H}_b,\vec{S}\right]\nn&&
\hspace{-15mm}~~=\vec{S}\times\left[\ga\vec{B}+2(1+\Phi)\left(\vec{\tilde{b}}+\frac{\bar{b}}{m}\hat{\vec{p}}
+\frac{\overrightarrow{(b_\mathrm{N})}}{m}\cdot\hat{\vec{p}}\right)\right]
-\frac{i}{m}\vec{S}\times\left[\bar{b}\vec{\nabla}\Phi+\overrightarrow{(b_\mathrm{N})}\cdot\vec{\nabla}\Phi\right],
\eea
where $\vec{\mu}\equiv\ga\vec{S}$ is the neutron magnetic moment and $\ga$ is the gyromagnetic ratio,
and we have defined $[\overrightarrow{(b_\mathrm{N})}\cdot\hat{\vec{p}}]^i\equiv(b_\mathrm{N})^i_{~k}\hat{p}_k$
and $[\overrightarrow{(b_\mathrm{N})}\cdot\hat{\vec{g}}]^i\equiv(b_\mathrm{T})^i_{~k}g^k$.
Now we investigate the correction to the Larmor frequency $\vec{\omega}_L=-\ga\vec{B}$. The tiny LV frequency correction
is dominated by $\de\vec{\omega}_L\equiv-2(1+\Phi)\vec{\tilde{b}}$
if assuming all the dimensional LV coefficients are roughly of the same order.
The $\left[\bar{b}~\hat{\vec{p}}+\overrightarrow{(b_\mathrm{N})}\cdot\hat{\vec{p}}\right]/m$ terms
are suppressed by $\sim10^{-11}$ for ultracold neutrons' velocity around $v\sim10$mm/s,
and the left terms in Eq.~(\ref{HeisenEqn}) are suppressed further by $|\vec{g}|/m\sim10^{-31}$,
so they can all be ignored with respect to the first $(1+\Phi)\vec{\tilde{b}}$ term.
As $\Phi=\vec{g}\cdot\vec{x}$, we also note that $\de\vec{\omega}_L$ depends on the height with respect to the altitude of the Laboratory.
This height-dependent CPT-odd couplings may be testable in more delicate atom interferometer experiments for polarized atoms
with large spatial separation \cite{DSO}.

To obtain the quantum Hamiltonian for the vertical motion, first we choose the laboratory coordinates in a way that
the positive $z$-direction is the upward vertical direction, and the positive $x,~y$ direction is along the
local south and east directions.
Then we adopt the ansatz in Ref.~\cite{Escobar1}:
assuming the horizontal motion is governed by the Gaussian wave-packet
$\psi(\vec{r}_\perp)\equiv\frac{1}{\sqrt{\pi}\si} \exp \left({i\vec{p}_\perp\cdot\vec{r}_\perp-\frac{\vec{r}_\perp^2}{2\si^2}} \right)$,
which is proper as the expectation value of the horizontal velocity is several orders of magnitude larger than the vertical one.
By averaging the full Hamiltonian $\hat{H}_\mathrm{full}=\hat{H}_\mathrm{LI}+(\hat{H}_\mathrm{LV})_B$ over the horizontal degree of freedom (d.o.f.),
we get an effective vertical Hamiltonian $\hat{H}_z
=\int{d^2\vec{r}_\perp}\psi^*(\vec{r}_\perp)\hat{H}_\mathrm{full}\psi(\vec{r}_\perp)=(\hat{H}_\mathrm{LI})_z+(\hat{H}_\mathrm{LV})_z$,
where
\bea\label{VerticalHLI}&&
\hspace{-20mm}(\hat{H}_\mathrm{LI})_z=m(1+gz)-\frac{ig}{2m}\hat{p}_z+\frac{1+gz}{2m}\left[\hat{p}_z^2+
\left(\sum_{a=1}^2p_a^2\right)+\frac{1}{\si^2}\right]
-\frac{1+gz}{4m}g(\vec{\si}\times\vec{p})_z,\\
\label{VerticalHLV}&&
\hspace{-20mm}(\hat{H}_\mathrm{LV})_z=-\vec{\si}\cdot\vec{\tilde{b}}(1+gz)-\frac{\bar{b}}{m}
\left[(1+gz)\left(\si_z\hat{p}_z+\sum_{a=1}^2\si_ap_a\right)-\frac{i}{2}\si_zg\right],
\eea
where $p_a$ with $a=x,~y$ are the classical horizontal momentum variable; for details, see \ref{VerticalO}.
As we want to discuss only the apparent rotation-invariant LV terms, the traceless $[b_\mathrm{T}]_k^i$ terms have already
been disregarded in Eq.~(\ref{VerticalHLV}).

\section{Review for a quantum particle in linear potential}\label{QMLinear}
For completeness, we briefly review the solution to the Schr$\ddot{\mathrm{o}}$dinger equation
\bea\label{BasicAiry}&&
\hspace{8mm}i\frac{\prt}{\prt t}\Psi(z,t)=\left[-\frac{1}{2m_I}\frac{\prt^2}{\prt z^2}+ m_Ggz\right]\Psi(z,t),
\eea
where $\Psi(z,t)=\int{dE}e^{-iEt}\rho(E)\phi_E(z)$, with $\phi_E(z)$ the solution to the stationary equation
$\phi''_E(z)+2m_IE\left(1-\frac{m_{_G}g}{E}z\right)\phi_E(z)=0$
and $\rho(E)$ the weighting factor for the Fourier expansion of the wave-packet $\Phi(z,t)$.
Note we work in natural unit such that $\hbar=c=1$, and we will insert back $\hbar$ and $c$ if necessary.
Defining the dimensionless variables $\tilde{z}=z/L_c$ and $\tilde{k}=kL_c,~\tilde{E}=E/E_c$
where $L_c\equiv\left(2m_{_I}m_{_G}g\right)^{-1/3}$ and $E_c\equiv\left[(m_{_G}g)^2/2m_{_I}\right]^{1/3}$ are the characteristic length and energy scales  respectively, we can cast the stationary equation into the dimensionless form
\bea\label{DSS}&&
\hspace{18mm}\phi''+(\tilde{E}-\tilde{z})\phi(\tilde{z})=0,
\eea
where for simplicity we have suppressed the subscript in $\phi_E(\tilde{z})$.
The solution to Eq.~(\ref{DSS}) is the famous Airy function
$\phi(\tilde{z})=c_1\mathrm{Ai}[\tilde{z}-\tilde{E}]$
as $\phi(\tilde{z})$ must be finite at $\tilde{z}\rightarrow+\infty$.
To fully determine the wave-function $\phi(\tilde{z})$, further requirement on boundary conditions must be imposed.
For example, for a trampoline or mirror in the qBounce experiment \cite{GRS},
which can be idealized as an infinitely high barrier and prohibits $\phi(\tilde{z})$ from penetrating into the $\tilde{z}<0$ zone,
we can impose $\phi(\tilde{z})|_{_{\tilde{z}=0}}=0$ to get a bouncer solution.

Denoting the zeros of $\mathrm{Ai}[-x]=0$ as $x_{n+1}$ where $n=0,1,2,\cdots$,
$\phi(\tilde{z})|_{_{\tilde{z}=0}}=0$ leads to a series of eigen-energies
\bea\label{BounceEi}&&
\hspace{-5mm}E_n=x_{n+1}E_c,\quad  ~~~~\{x_1, x_2, x_3, \cdots\}=\{2.338, 4.088,
5.521, \cdots\},
\eea
where we intentionally offset $n$ by 1 to make $E_0$ the ground state eigen-energy.
The eigen-function is given by
\bea\label{BounceEf}
\hspace{8mm}\phi_n(z)=\mathrm{Ai}\left[\frac{z}{L_c}-x_{n+1}\right]/\left(L_c^{\hf}\left|\mathrm{Ai}'[-x_{n+1}]\right|\right).
\eea
For the neutron bouncing problem, $L_c=5.871\mu$m.
We keep distinctive notations between the inertial mass $m_{_I}$ and the gravitational mass $m_{_G}$
to emphasis that the quantum test of equivalence principle with neutron experiment can be quite promising \cite{IGMQM}.
Note $z_{n+1}=x_{n+1}L_c$ corresponds to the classical turning height for $\phi_n(z)$.

In contrast to the classical case, the turning height $z_{n+1}$ is also quantized \cite{UCNILL},
a natural consequence of the energy quantization condition.
However, largely due to the extreme weakness of gravity, $g=9.818\mathrm{m/s}^2=2.156\times10^{-32}\mathrm{GeV}$,
the characteristic energy is $E_c=0.601$peV. To detect this quantization is quite difficult, as the relevant neutron energy
also needs be around the peV level. Indeed, with sophisticated experimental designs and ultracold neutron (UCN) beam,
ILL has already confirmed the discrete nature of neutrons' gravitational bound states \cite{UCNILL}.
In the following we consider the spin-dependent LV corrections in Eq.~(\ref{VerticalHLV}) to the bound states.

\section{Spin-dependent LV corrections to the gravitational bound state}\label{QMLVb}
To analyze the LV corrections, first we give a rough estimate of the energy scales
for various operators with the parameters in Ref.~\cite{GRS}.
The estimate on the energy scales for various operators in Eqs.~(\ref{VerticalHLI}) and (\ref{VerticalHLV}) are shown in Table \ref{EEO}.
From Table \ref{EEO}, it is clear that the energy budget is horizontal motion dominating, such as
$\frac{p_a^2}{2m}gz\sim2.1\times10^{-16}$peV$\gg\langle gz\frac{p_z^2}{2m}\rangle\sim10^{-21}$peV.
However, since the horizontal d.o.f. has already been averaged out and may only contribute to an irrelevant phase factor
supposing $\sum_{a=x,y}p_a^2/(2m)\simeq0.21\mu$eV ($|\vec{v}_{\perp}|\simeq7$m/s),
we will ignore them in the following discussions and concentrate to the operators with the vertical d.o.f. only.
The leading order LI contribution comes from the first two terms in Eq.~(\ref{VerticalHLI}), while all the other
sub-leading-order LI corrections can be distinguished by the direction-independent nature from the LV corrections.
The latter can be obtained through the LV signals with characteristic sidereal frequency.
So in the following we will only keep the first two LI operators.
For the LV operators, from the rough estimate, we ignore $i\bar{b}g\si_z/(2m)$ and $-\bar{b}gz\si_ap_a/m$, corresponding to the last
two terms in Table \ref{EEO}, while for completeness, we also consider the second term, $gz\vec{\si}\cdot\vec{\tilde{b}}$.
In summary, we ignore all terms with expectation values less than $10^{-15}$peV in the rough estimate except for the $\tilde{b}$-term.
\begin{table}
 \centering
\caption{\label{EEO}A rough estimate of the energy budget for the expectation values of various LI and LV operators.
   The data of the UCN mean horizontal velocity are from Ref.~\cite{GRS}, while for the vertical motion, we roughly assume
   $mgz\sim\hat{p}_z^2/(2m_{_I})\sim$peV. For the estimate of LV operators, we choose the conservative bounds from
   either the UCN experiment \cite{SPUCN} or the Cs spectroscopy \cite{CsSpec} on the corresponding LV coefficients.}
  \footnotesize
\begin{center}
\begin{tabular}{@{}llll}
\br
LI Operators          &    Energy Estimate (peV)                 & LV Operators                                 &    Energy Estimate(peV)\\
\mr
$\langle mgz\rangle$                   & ~~~1                   &$\langle\vec{\si}\cdot\vec{\tilde{b}}\rangle$ & $~~~\leq10^{-8}$\cite{SPUCN}\\
$\langle\frac{\hat{p}_z^2}{2m}\rangle$ & ~~~1                   &$\langle gz\vec{\si}\cdot\vec{\tilde{b}}\rangle$
&~~~$\leq10^{-29}$\cite{SPUCN}\\
$\sum_{a=x,y}\frac{p_a^2}{2m}$         &~~~$2.108\times10^5$  &$\bar{b}/m\langle(\si_z\hat{p}_z)\rangle$     &~~~$\leq10^3$\cite{CsSpec}\\
$\langle\frac{g\hat{p}_z}{2m}\rangle$  &~~~$10^{-21}$         &$\bar{b}/m\langle\si_ap_a\rangle$             &~~~$\leq10^6$\cite{CsSpec}\\
$\langle gz\frac{p_z^2}{2m}\rangle$    &~~~$10^{-21}$         &$\bar{b}/m\langle g\si_z\rangle$              &~~~$\leq10^{-17}$\cite{CsSpec}\\
$\langle(\vec{\si}\times\vec{p})_z\frac{g}{4m}\rangle$ &~~~$10^{-18}$   &$\bar{b}/m\langle gz\si_ap_a\rangle$ &~~~$\leq10^{-15}$\cite{CsSpec}\\
$\langle gz\rangle\left(\sum_{a=x,y}\frac{p_a^2}{2m}\right)$          &~~~$2.1\times10^{-16}$  &$\bar{b}\langle(\si_z\hat{p}_z)gz/m\rangle$
&~~~$\leq10^{-18}$\cite{CsSpec}\\
\br
\end{tabular}
\end{center}

\end{table}

\normalsize

In other words, we only consider the LV correction
\bea\label{EffLVH}
(\hat{H}_\mathrm{LV})_z\supset-\vec{\si}\cdot\vec{\tilde{b}}(1+gz)-\frac{\bar{b}}{m}\left(\si_z\hat{p}_z+\sum_{a=1}^2\si_ap_a\right).
\eea
The leading order perturbation in Eq.~(\ref{EffLVH}) is $-\vec{\si}\cdot\vec{\tilde{b}}$, where
$\vec{\tilde{b}}$ behaves like an effective magnetic field and can mimic either neutron magnetic interaction $-\vec{\mu}\cdot\vec{B}$,
or the spin-rotation coupling $\vec{\si}\cdot\vec{\Omega}$.
With an idealized multi-layer magnetic screen or by intentionally alternating the magnetic field direction,
the orthogonal component of $\vec{\tilde{b}}$ to the Earth's rotation axis has been constrained by clock comparison experiment at least
to $|\vec{\tilde{b}}_\perp|\leq10^{-29}$GeV \cite{SPUCN}. A more recent comagnetometer experiment based on the detection of
freely precessing nuclear spins from polarized $^3\mathrm{He}$ and $^{129}\mathrm{Xe}$ gas samples with SQUID detectors
has improved the bounds to $|\vec{\tilde{b}}_\perp|\leq8.4\times10^{-34}$GeV \cite{ComagSQUID}.
The ${b}^0\subset\bar{b}$ has been constrained by the Cs spectroscopy to
the level $b^0\leq3\times10^{-7}$GeV \cite{CsSpec}.
As both $\vec{\tilde{b}}$ and $\bar{b}$ only have upper bounds up to now, we assume that they are roughly of the same order and
treat the corresponding operators as small perturbations to the LI Hamiltonian
$\hat{H}_0\equiv-\frac{1}{2m_I}\frac{\prt^2}{\prt z^2}+ m_Ggz$ from now on.

In principle, $\tilde{b}^i$ can be sidereal time-dependent and solar time-dependent due to the Earth's rotational
and orbital motions respectively.
However, since the sidereal period is much larger than the characteristic time scale for the GRS experiment, 
$T_{\oplus}=23.56\mathrm{h}\gg\tau\simeq23\mathrm{ms}$ \cite{GRS}, and the solar period $\sim1$ year is even larger,
the time variation of $\vec{\tilde{b}}$ is irrelevant in the simple analysis of the GRS experiment.
However, as we want our analysis to be applicable to other experiments, including the continuous precise measurement of the
gravitational transition frequency for GRS experiment in the future,
we will consider the time-independent $\tilde{b}^i$ first, and then assume $\vec{\tilde{b}}(t)\equiv{}B_0\left(\sin\th\cos[\omega{t}+\phi],\sin\th\sin[\omega{t}+\phi],\cos\th\right)$
as a preliminary time-dependent case study. Finally, we will take the more realistic sidereal time dependence into account.
Also we note that the simple $\vec{\tilde{b}}(t)$ case can be viewed as a consequence of
putting the whole apparatus on a turntable with adjustable rotation frequency $\omega$ and
$\omega\gg\omega_{\oplus}=2\pi/T_{\oplus}$.

\subsection{The $-\vec{\si}\cdot\vec{\tilde{b}}$ correction}

For simplicity, let us consider a constant $\vec{\tilde{b}}\equiv{}B_0\left(\sin\th\cos\phi,\sin\th\sin\phi,\cos\th\right)$ first,
and discuss the time-dependent $\vec{\tilde{b}}(t)$ later.

\subsubsection{The simple case with a time dependent $\vec{\tilde{b}}$}\label{timeDH}

\

\noindent
Note that the solution to the Schr$\ddot{\mathrm{o}}$dinger equation,
$i\dot{\Psi}=[\hat{H}_0-B_0\vec{\si}\cdot\hat{n}]\Psi$, is separable,
and the question is reminiscent of the 2-fold degenerate energy levels in a constant magnetic field.
So there are two towers of eigenstates
\bea\label{NaiveSol1}&&
\xi_n=\left(
                     \begin{array}{c}
                      \cos(\th/2) \\
                       \sin(\th/2)e^{i\phi} \\
                     \end{array}
                   \right)
\phi_n(z)e^{-i(E_n-B_0)t},\\&&\label{NaiveSol2}
\eta_n=\left(
                     \begin{array}{c}
                      \sin(\th/2)e^{-i\phi} \\
                      -\cos(\th/2) \\
                     \end{array}
                   \right)\phi_n(z)e^{-i(E_n+B_0)t},
\eea
where $\phi_n(z)$
is given by Eq.~(\ref{BounceEf}) and $E_n$ ($n=0,1,2,\cdots$) is given by Eq.~(\ref{BounceEi}).
A general solution is a superposition,
$\Psi(t,z)=\sum_n[c_{1n}\xi_n+c_{2n}\eta_n]$,
where $c_{1n},~c_{2n}$ are two series of constants subject to initial and normalization conditions.
For a pair of nonvanishing $c_{1n},~c_{2m}$ with $n\neq{m}$,
an entangled state between spin and vertical motion can be constructed due to this ``cosmic spin anisotropy field"
\cite{SPUCN}.

Now consider the case of the simple time dependence
$\vec{\tilde{b}}(t)=B_0\left(\sin\th\cos[\omega{t}+\phi],\right.\\
\left.\sin\th\sin[\omega{t}+\phi],\cos\th\right)$.
The time-dependent Hamiltonian is $\hat{H}_t=\hat{H}_0-\vec{\si}\cdot\vec{\tilde{b}}(t)$,
and thus has two towers of instantaneous ``eigenstates" $\psi_{n\pm}(t)$
with the same form as Eqs.~(\ref{NaiveSol1}) and (\ref{NaiveSol2}), except for the replacement of $\phi$ by $\phi+\omega{t}$.
Then we can promote $c_{1n},~c_{2n}$ in $\Psi=\sum_n[c_{1n}\psi_{n+}+c_{2n}\psi_{n-}]$
to be time-dependent to find analytical solutions.
Finally, we get
\bea\label{ansatz1}
\hspace{8mm}\Psi(t,z)=\sum_n\left(
       \begin{array}{c}
         f_{1}(t)e^{-i\frac{\omega}{2}t} \\
         f_{2}(t)e^{i(\phi+\frac{\omega}{2}t)} \\
       \end{array}
     \right)\phi_ne^{-iE_nt},
\eea
where
\bea\label{TimDeCoe1}&&
\hspace{-20mm}f_{1}(t)=\left[d_{1n}e^{-i\Omega{t}}+d_{2n}e^{+i\Omega{t}}\right],\\
\label{TimDeCoe2}&&
\hspace{-20mm}f_{2}(t)=\frac{-1}{B_0\sin\th}\left[d_{1n}\left(\frac{\omega}{2}+B_0\cos\th+\Omega\right)e^{-i\Omega{t}}
+d_{2n}\left(\frac{\omega}{2}+B_0\cos\th-\Omega\right)e^{+i\Omega{t}}\right],
\eea
and $\Omega\equiv\sqrt{\frac{\omega^2}{4}+\omega B_0\cos\th+B_0^2}$.
Note that $d_{1n},~d_{2n}$ are just two series of constants marked with $n$ but do not necessarily depend on $n$.
They can be determined by the initial and normalization conditions $\left(\int_{0}^{+\infty}dz|\Psi(0,z)|^2=1\right)$.
We emphasize that Eq.~(\ref{ansatz1}) is not the superposition of eigen-states with specific $n$,
though superficially it looks like so, as indicated by the explicit time dependence of $\hat{H}_t$.
In other words, Eq.~(\ref{ansatz1}) is the most general wave solution,
while for some specific solutions, only a few constants $d_{1n},~d_{2n}$ are needed to be nonzero.
For example, we can freely choose $d_{1n},~d_{2n}$ to construct an entangled state
{\footnotesize\bea\label{entangleS1}&&
\hspace{-25mm}\Psi_{n\neq{m}}=\left(
                     \begin{array}{c}
                       \frac{B_0\sin\th}{2\Omega}e^{-i\frac{\omega}{2}t} \\
                       -\frac{\frac{\omega}{2}+B_0\cos\th+\Omega}{2\Omega}
                       e^{i[\phi+\frac{\omega}{2}]t} \\
                     \end{array}
                   \right)\phi_n(z)e^{-i[E_n+\Omega]t}
+\left(
       \begin{array}{c}
        -\frac{B_0\sin\th}{2\Omega}e^{-i\frac{\omega}{2}t} \\
        \frac{\frac{\omega}{2}+B_0\cos\th-\Omega}{2\Omega}
                       e^{i[\phi+\frac{\omega}{2}]t} \\
                     \end{array}
                   \right)\phi_m(z)e^{-i[E_m-\Omega]t}.\nn
\eea
}%
The entanglement between spin and the vertical motion due to the LV spin coupling may lead to interesting
phenomena for polarized neutrons, and proper manipulation of entangled state may enable
high precision probes of LV for neutral atom experiments \cite{YouliLV}.

A more simpler example is an initially spin-up state, whose
time evolution is given by
{\small
\bea\label{SpinUp2}&&
\Psi_\uparrow(t,z)=
            \left(
              \begin{array}{c}
                \left[\cos(\Omega{t})+i\frac{(\frac{\omega}{2}+B_0\cos\th)}{\Omega}\sin(\Omega{t})\right]
                e^{-i\frac{\omega}{2}t} \\
                \frac{iB_0\sin\th}{\Omega}\sin(\Omega{t})e^{i[\frac{\omega}{2}t+\phi]} \\
              \end{array}
            \right)\phi_n(z)e^{-iE_nt},\nn
\eea}%
While for a horizontally polarized initial state such as an eigenstate of $\si_x$, the time evolved state is
{\small
\bea\label{SpinHorizT}&&
\hspace{-15mm}\Psi_\perp(t,z)=\frac{1}{\sqrt{2}}\left(
              \begin{array}{c}
                \left[\cos(\Omega{t})+i\sin(\Omega{t})
                \left(\frac{\frac{\omega}{2}+B_0(\cos\th+\sin\th{e^{-i\phi}})}{\Omega}\right)\right]e^{-\frac{i\omega}{2}t} \\
                \left[\cos(\Omega{t})-i\sin(\Omega{t})
                \left(\frac{\frac{\omega}{2}+B_0(\cos\th-\sin\th{e^{i\phi}})}{\Omega}\right)\right]e^{+\frac{i\omega}{2}t}\\
              \end{array}
            \right)\phi_n(z)e^{-iE_nt}.
\eea
}

We can define the polarization asymmetry as the probability difference in finding the particle in the spin-up and spin-down states,
$\frac{\mathrm{P}_{\uparrow}-\mathrm{P}_{\downarrow}}{\mathrm{P}_{\uparrow}+\mathrm{P}_{\downarrow}}$.
Then we obtain from Eq.~(\ref{SpinUp2}) the polarization asymmetry
\bea\label{PolarUp}&&
\left(\frac{\mathrm{P}_{\uparrow}-\mathrm{P}_{\downarrow}}{\mathrm{P}_{\uparrow}+\mathrm{P}_{\downarrow}}\right)[\Psi_\uparrow]
=1-2\frac{B_0^2 \sin^2\th}{\Omega^2}\sin^2[\Omega t].
\eea
Clearly, the deviation of the asymmetry from 1 for initial spin-up state
is proportional to $\mathcal{O}(B_0^2)$, which is very difficult to probe
considering that the magnitude of LV coefficients must be very tiny.
For the state (\ref{SpinHorizT}), the asymmetry is
\bea\label{SpinDif1}&&
\hspace{-15mm} \left(\frac{\mathrm{P}_{\uparrow}-\mathrm{P}_{\downarrow}}{\mathrm{P}_{\uparrow}+\mathrm{P}_{\downarrow}}\right)[\Psi_\perp]
=\frac{B_0\sin\th}{\Omega}\left[\sin(2\Omega{t})\sin\phi+\frac{\omega+2B_0\cos\th}{\Omega}\sin^2(\Omega{t})\cos\phi\right],
\eea
which is of $\mathcal{O}(B_0)$, hence the linear order of $|\vec{\tilde{b}}|$. Therefore the state (\ref{SpinHorizT}) will be more sensitive to the test of spin-dependent LV effects.
\begin{figure}
\centering
 \subfigure[~$\theta_\mathrm{Lab}$ as a function of time $T$ and $\th_\odot$]{\label{thetaLab}\includegraphics[width=68mm]{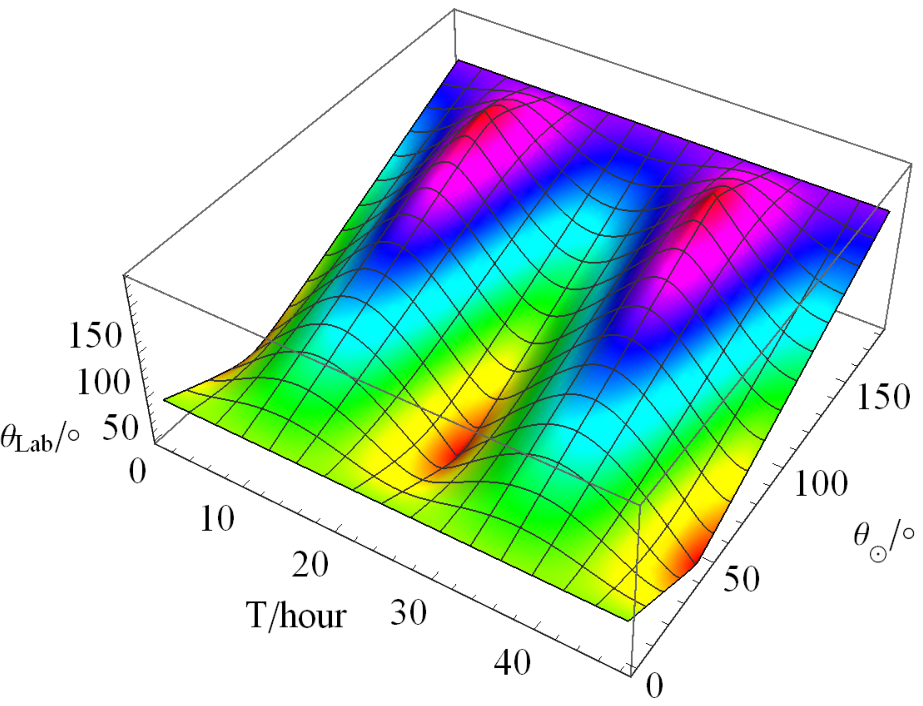}}
  \hspace{0.2in}
 \subfigure[~Asymmetry as a function of time $t$ and $\omega$]{\label{Asymmetry}\includegraphics[width=70mm]{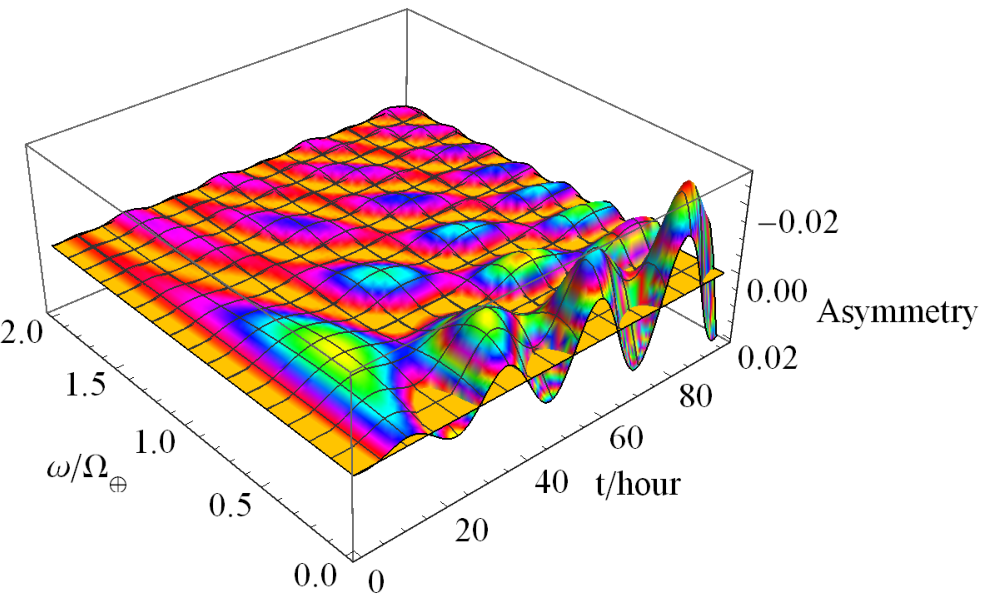}}
       \caption{\footnotesize The unit of time in both figures is hour, and the Earth rotation frequency is $\Omega_\oplus\simeq2\pi/(23\mathrm{h}56\mathrm{min})$.
        (a) $\theta_\mathrm{Lab}$ as a function of time $T$ and $\th_\odot$ in Sun-centered frame with $\phi_\odot=0$.
        The angles are in unit of degree. $\th_\odot$ denotes the $\th$ angle in the Sun-centered frame.
        We choose the colatitude of ILL in Grenoble, $\chi=45.2066^\circ$ \cite{ILLGeo}. 
        (b) Asymmetry (\ref{SpinDif1}) as a function of time $t$ and artificially introduced rotation frequency $\omega$.
        For convenience, we choose the conservative bound for the LV coefficient, $B_0=|\vec{\tilde{b}}|\simeq3.7\times10^{-32}$GeV.
        The rotation frequency $\omega$ is plot in unit of $\Omega_\oplus$.
         }\label{SiderealLab}
\end{figure}

\subsubsection{Transforming to the Sun-centered Frame}

\

\noindent
In the above discussions, the $\tilde{b}$-term which has indices in the lowercase denotes the LV coefficients in the laboratory frame.
Due to the Earth's rotation, these coefficients have a sidereal time-dependence, and can be related to the LV counterparts with indices
denoted by the capital letters in the Sun-centered frame by a rotation \cite{AKCL}\cite{2002Sun}
{\footnotesize
\bea\label{rotaSid}&&
     \left(
       \begin{array}{c}
        \tilde{b}_x \\
        \tilde{b}_y \\
        \tilde{b}_z \\
       \end{array}
     \right)
     =\left(
              \begin{array}{c}
                 \cos(\chi)\left(\tilde{b}_X\cos[\Omega_\oplus{T}]+\tilde{b}_Y\sin[\Omega_\oplus{T}]\right)-\tilde{b}_Z\sin\chi \\
                 \tilde{b}_Y\cos[\Omega_\oplus{T}]-\tilde{b}_X\sin[\Omega_\oplus{T}] \\
                 \sin(\chi)\left(\tilde{b}_X\cos[\Omega_\oplus{T}]+\tilde{b}_Y\sin[\Omega_\oplus{T}]\right)+\tilde{b}_Z\cos\chi \\
               \end{array}
             \right),\nonumber\\
 \eea
 }%
where $\chi$ is the colatitude of the laboratory.
The Sun-centered frame is the standard approximate inertial frame where LV coefficients are assumed to be constants.
The angles $\th,~\phi$ in the laboratory frame also have to be replaced by
\bea\label{PhiL1}&&
\hspace{-20mm}\phi_\mathrm{Lab}
 \equiv\tan^{-1}\left[\frac{\tilde{b}_y}{\tilde{b}_x}\right]=
 \cot^{-1}\left[\cos\chi\cot(\phi_\odot-\Omega_\oplus{T})-\sin\chi\cot[\th_\odot]
\csc(\phi_\odot-\Omega_\oplus{T})\right],\\
\label{ThetaL1}&&
\hspace{-20mm}\theta_\mathrm{Lab}
 \equiv\cos^{-1}\left[\frac{\tilde{b}_z}{|\tilde{b}|}\right]
 =\cos^{-1}\left[\cos\chi\cos\th_\odot+\sin\chi\sin\th_\odot\cos(\phi_\odot-\Omega_\oplus{T})\right],
\eea
where we explicitly denote the laboratory angles by $\theta_\mathrm{Lab}$ and $\phi_\mathrm{Lab}$
to emphasis the time dependence, and $\th_\odot,~\phi_\odot$ are the angles of $\vec{\tilde{b}}$
in the Sun-centered frame, \ie, $\vec{\tilde{b}}=B_0(\sin\th_\odot\cos\phi_\odot,\sin\th_\odot\sin\phi_\odot,\cos\th_\odot)$.
The parameter $\Omega_\oplus\simeq2\pi/(23\mathrm{h}56\mathrm{min})$ is the Earth's rotation frequency,
$\chi=45.2066^\circ$ is the colatitude of the laboratory in ILL in Grenoble \cite{ILLGeo}.
With the standard convention \cite{DATA}, $T$ is the time measured in the Sun-centered frame from the
time when $y$ and $Y$ axes coincide, and is chosen for convenience for each experiment \cite{2002Sun}.

Since $\phi_\odot$ can be assigned any value in the interval $[0,\pi)$ by a proper coordinate choice (a proper choice of the
starting point of $T$),
we show the time-dependence of $\theta_\mathrm{Lab}$ as a function of $\th_\odot$ and $T$
with $\phi_\odot=0^\circ$ in Fig.\ref{thetaLab}. We do not demonstrate $\phi_\mathrm{Lab}$
with $\theta_\mathrm{Lab}$ in Fig.\ref{SiderealLab} due to two reasons. First,
the $\cos x$ function is a monotonic function for $x\in[0,\pi)$, while $\cot x$ (or $\tan x$)
is a multi-valued function for $x\in[0,2\pi)$. Therefore directly plotting $\phi_\mathrm{Lab}$ in Eq.~(\ref{PhiL1})
will show discontinuities. Second, $\theta_\mathrm{Lab}$ already demonstrates the sidereal variation,
which can be easily seen from the sine-like color pattern shown in Fig.\ref{thetaLab}.
In Fig.\ref{SiderealLab} we also plot the asymmetry (\ref{SpinDif1}) as a function of time $t$
and the artificially introduced rotation frequency $\omega$ with $|\vec{\tilde{b}}|\simeq3.7\times10^{-32}$GeV,
a conservative bound from Ref.~\cite{3He-129Xe}.
Note in plotting Fig.\ref{Asymmetry}, we have already substituted Eq.~(\ref{rotaSid}) into Eq.~(\ref{SpinDif1}).
Clearly, we see that the time evolution of the asymmetry has approximately a 24-hour period,
a manifestation of the sidereal effect. More interestingly, the asymmetry accumulates with increasing time,
showing the time-increased depolarization due to LV for sufficiently stable polarized neutrons.
This may be a new way to probe the spin-dependent LV effects.
\begin{figure*}
\centering
 \subfigure[~The probability in $\Psi_\perp$ state with $\omega=0$]{\includegraphics[width=70mm]{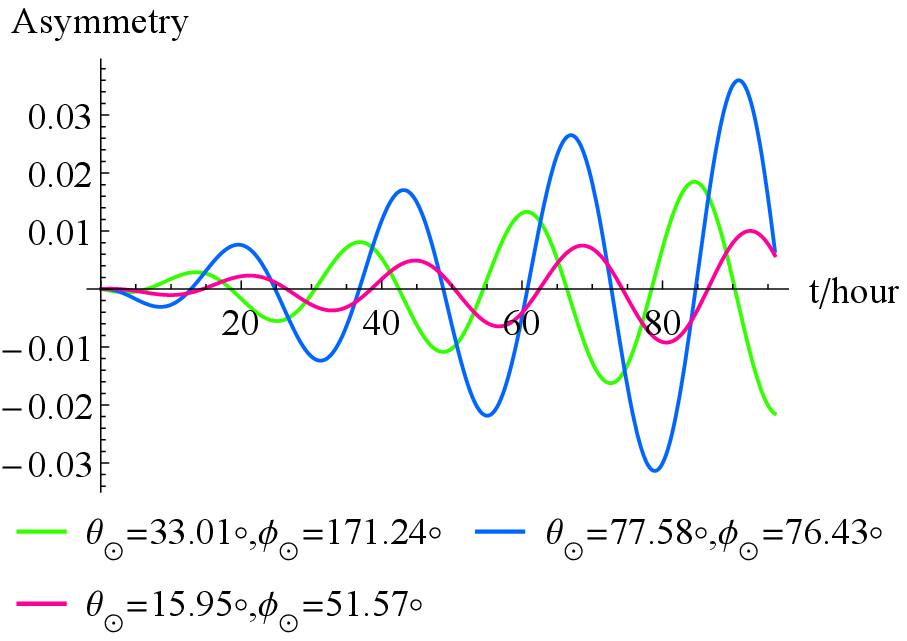}\label{ProbX}}
  \hspace{0.2in}
 \subfigure[~The trace of the tip of the spin]{\includegraphics[width=60mm]{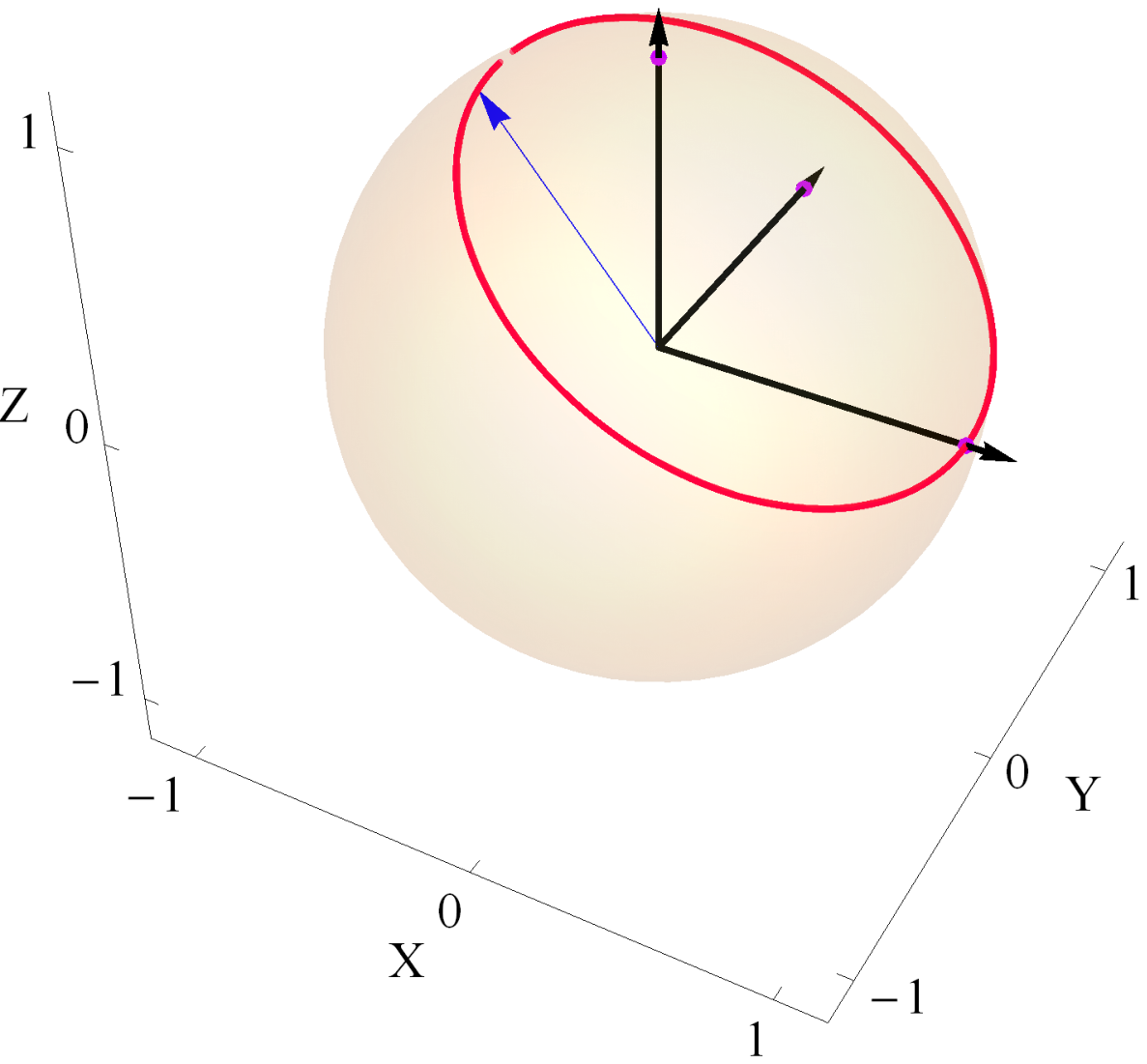}\label{SpinX}}
       \caption{\footnotesize
        (a) The asymmetry for different positions in the Sun-centered frame. We choose $B_0=|\vec{\tilde{b}}|\simeq3.7\times10^{-32}$GeV \cite{3He-129Xe}, and a random set of angles
        $(\th_\odot,\phi_\odot)$ shown in the legends of the figure.
        (b) The trace of the tip of the polarization vector for an initial state with spin pointing towards the X-direction.
        We choose the same $B_0=\pi/(120\mathrm{ms})$, and the other parameters are $\th=\pi/6,~\phi=0,~\omega=0$.
        The time for the evolved blue arrow is chosen as $t=176$ms, which is an order larger than the characteristic
        time ($15\sim23$ms) in the neutron experiment in ILL \cite{GRS}.}
\end{figure*}

In Fig.\ref{ProbX}, we plot the time dependence of the asymmetry (\ref{SpinDif1})
with ${\omega=0}$,
\bea\label{asymSid}&&
\hspace{-15mm}\frac{\mathrm{P}_{\uparrow}-\mathrm{P}_{\downarrow}}{\mathrm{P}_{\uparrow}+\mathrm{P}_{\downarrow}}[\Psi_\perp]
=2\sin\th\sin(B_0t)\left[\cos(B_0t)\sin\phi+\cos\th\sin(B_0t)\cos\phi\right]\simeq2t\tilde{b}_y\nn&&~~~~~~
=2t\left[\tilde{b}_Y\cos(\Omega_\oplus t)-\tilde{b}_X\sin(\Omega_\oplus t)\right],
\eea
where we have used $\sin(B_0t)\simeq B_0t$ and kept terms only at the linear order of $B_0$ at the second
approximation. In the last equality we have substituted the Sun-centered frame $\tilde{b}$-component.
The above approximation also means $B_0t\ll1$. For the LV magnitude $|\tilde{b}|$ still chosen as $3.7\times10^{-32}\mathrm{GeV}\sim7.7\times10^{-4}\Omega_\oplus$, time range must satisfy $t\leq10/\Omega_\oplus\simeq10$days.
In Fig.\ref{ProbX}, a set of randomly chosen angles $(\th_\odot,\phi_\odot)$ were used such that
$(b_X,b_Y,b_Z)=B_0(\sin\th_\odot\cos_\odot,\sin\th_\odot\sin_\odot,\cos\th_\odot)$.
The time increasing feature and the sidereal time dependence are also manifested in Fig.\ref{ProbX}. The amplitude
of the asymmetry is larger for larger $\th_\odot$, reflecting the fact that the asymmetry depends heavily on $b_X,~b_Y$ rather than
$b_Z$. For complementarity, we also demonstrate the spin precession on the Bloch sphere in Fig.\ref{SpinX}.
In fact, the asymmetry is a manifestation of the spin precession, and the latter can be measured very accurately
by comagnetometer \cite{comagnetometer}, or by the comparison of the ratio of Zeeman level frequency difference by
reversing the reference magnetic field \cite{silenko}.

\subsection{The $(\vec{\si}\cdot\vec{\tilde{b}})gz$ and $\frac{\bar{b}}{m_I}\si_z\hat{p}_z$ corrections}\label{LVQT}

In this subsection, we investigate the $(\vec{\si}\cdot\vec{\tilde{b}})gz$ and $\frac{\bar{b}}{m_I}\si_z\hat{p}_z$ corrections.

\subsubsection{The $-\vec{\si}\cdot\vec{\tilde{b}}(1+gz)$ correction}

\

\noindent
As $\bar{b}$ and $\vec{\tilde{b}}$ contain different LV components by definition,
and $\frac{\bar{b}}{m_I}\si_z\hat{p}_z$ is suppressed by a factor $\frac{\langle\hat{p}_z\rangle}{m_I}\sim\frac{v_z}{c}\simeq10^{-11}$
compared with $\vec{\si}\cdot\vec{\tilde{b}}$ (assuming $\tilde{b}$ and $\bar{b}$ are roughly the same order),
we consider them separately.
Currently we consider the correction $(\vec{\si}\cdot\vec{\tilde{b}})gz$ first, and
the Hamiltonian $\hat{H}_{\tilde{b}}=\hat{H}_0-\vec{\si}\cdot\vec{\tilde{b}}(1+gz)$ then
contains a coupling between spin and vertical variables.

Unlike in the previous cases, the spin and vertical motion is inseparable
for the Hamiltonian $\hat{H}_{\tilde{b}}$.
Still assuming $\vec{\tilde{b}}\equiv{}B_0(\sin\th\cos\phi,\sin\th\sin\phi,\cos\th)$,
the two towers of eigen-solutions are given by
\bea\label{NaiveSol}&&
\xi_{n+}=\left(
                     \begin{array}{c}
                      \cos(\th/2) \\
                       \sin(\th/2)e^{i\phi} \\
                     \end{array}
                   \right)e^{-i(E_{n+}-B_0)t}\phi_{n+}(t,z),\nn &&
\eta_{n-}=\left(
                     \begin{array}{c}
                      \sin(\th/2)e^{-i\phi} \\
                      -\cos(\th/2) \\
                     \end{array}
                   \right)e^{-i(E_{n-}+B_0)t}\phi_{n-}(t,z),\nonumber
\eea
where $L_{c\pm}\equiv \left(2m_{_I}m_{_{G\mp}}g\right)^{-1/3}$,
$E_{n\pm}\equiv \left[\frac{(m_{_{G\mp}}g)^2}{2m_{_I}}\right]^{1/3}x_{n+1}$, and
$\phi_{n\pm}(z)\equiv\frac{\mathrm{Ai}\left[z/L_{c\pm}-x_{n+1}\right]}{L_{c\pm}^{\hf}\mathrm{Ai}'\left[-x_{n+1}\right]|}$.
Note that $m_{_{G\mp}}\equiv m_{_G}\mp B_0$ can be viewed as the spin-dependent gravitational mass
induced by the LV $(\vec{\si}\cdot\vec{\tilde{b}})gz$ term.
This may be detectable in the test of weak equivalence principle with neutrons \cite{Tino2014}\cite{NeutDiff}.
Since
\bea\label{ExactEn}&&
E_{n\pm}\equiv \left[\frac{\left(m_{_{G\mp}}g\right)^2}{2m_I}\right]^{1/3}x_{n+1}
\simeq
x_{n+1}E_c\left[1\mp\frac{2B_0}{3m_{_G}}\right],
\eea
the eigen-energies $E_{n\pm}\mp B_0$ can lead to detectable LV corrections to
the transition frequency between different gravitational states,
which will be discussed in detail in Sec.~\ref{Perturb}.
A general state is a superposition $\Psi=\sum_n[c_{1n}\xi_{n+}+c_{2n}\eta_{n-}]$,
where $c_{1n},~c_{2n}$ are two sets of constants subject to initial
and normalization conditions $\sum_n|c_{1n}|^2+|c_{2n}|^2=1$.
As the external motion (encoded in $\phi_{n\pm}(t,z)$) is closely intertwined with the spin d.o.f.
(encoded in $\xi_{n+}$ and $\eta_{n-}$), in general it is impossible to prepare an arbitrarily polarized initial state.
Also note that the LV spin-gravity coupling is suppressed by a factor $2E_c/(3m_{_G})\sim10^{-21}$ for peV neutrons,
only very dedicate interferometer experiment may have the potential to probe it.

\subsubsection{The $-\si_z\hat{p}_z$ correction}\label{SpinMoment}

\

\noindent
Since $\langle\hat{p}_z\rangle/p_a\sim10^{-3}$ and the horizontal $p_a$ can be
treated as classical variables, we may combine $\frac{\bar{b}}{m}p_a+\tilde{b}_a$ together
and consider only the $\si_z\hat{p}_z$ term in the Hamiltonian below,
\bea\label{MomentumH}&&
\hat{H}_{\bar{b}}=\hat{H}_0-\frac{\bar{b}}{m_I}\si_z\hat{p}_z
=\frac{1}{2m_I}\left[\hat{p}_z-\bar{b}\si_z\right]^2+m_{_G}gz-\frac{\bar{b}^2}{2m_I}.
\eea
Hamiltonian (\ref{MomentumH}) is simply $\hat{H}_0-\frac{\bar{b}^2}{2m_I}$
with a momentum shift $\hat{p}_z\rightarrow\hat{p}_z-\si_z\bar{b}$,
so the eigen-solution can be obtained by applying the momentum shifting operator $e^{i\si_z\bar{b}'z}$
to the Airy function.
The general solution to $\hat{H}_{\bar{b}}$ is
\bea\label{Spin-Momentum}
\Psi(t,z)=\sum_{n}\left(
              \begin{array}{c}
                c_{1n}~e^{i\bar{b}z} \\
                c_{2n}~e^{-i\bar{b}z}\\
              \end{array}
            \right)\phi_n(z)e^{-i\left(E_n-\frac{\bar{b}^2}{2m_I}\right)t}.
\eea
Up to a phase choice, the set of constants $c_{1n},~c_{2n}$ are subject to initial condition and the
normalization condition $\sum_n|c_{1n}|^2+|c_{2n}|^2=1$.
Note that the spin-momentum coupling $\si_zp_z$ ensures the close bound between spin and the vertical motion.

As a concrete example, let us consider
\bea\label{HorizLike}
\Psi_\mathrm{n}=\frac{1}{\sqrt{2}}\left(
                                    \begin{array}{c}
                                      e^{i\bar{b}z} \\
                                      e^{-i\bar{b}z} \\
                                    \end{array}
                                  \right)\phi_n(z)e^{-i\left(E_n-\frac{\bar{b}^2}{2m_I}\right)t}.
\eea
Note that $\Psi_\mathrm{n}$ is not a horizontal spin-eigenstate,
though $\langle\Psi_\mathrm{n}|\si_z|\Psi_\mathrm{n}\rangle=0$ and
detecting neutron in spin-up or spin-down state will both have $50\%$ probability.
Taking into account of the fact that $\bar{b}z\ll1$ for $z\sim10^{-3}$m, we have
\bea\label{SpinVEV}&&
\langle\Psi_\mathrm{n}|\si_x|\Psi_\mathrm{n}\rangle\simeq1-\mathcal{O}(\bar{b}^2),\quad
\langle\Psi_\mathrm{n}|\si_y|\Psi_\mathrm{n}\rangle\simeq-\frac{4}{3}\bar{b}L_cx_{n+1}.
\eea
Interestingly, state $\Psi_\mathrm{n}$ is very close to the spin-$x$ polarized state,
however, its expectation value for spin-$y$ polarization depends on the motional energy,
indicated by the dependence on $x_{n+1}$. In other words, $\Psi_\mathrm{n}$ and  $\Psi_\mathrm{m}$
for $n\neq m$ belong to states with slightly different spin-polarization,
and $\Psi_\mathrm{n}$ has a vertical distribution for horizontal spin detection,
a peculiar feature of spin-momentum coupling.

More analytical models are also possible.
For example, keeping the much tiny $gz\si_z\hat{p}_z$ term in $\hat{H}_{\bar{b}}$ can also lead to analytical solutions
with Hermite and confluent Hypergeometric functions involved, and therefore are more difficult to deal with.
We leave them to future work.

\subsubsection{Perturbative analysis with both $\vec{\tilde{b}}$- and $\bar{b}$-terms}
\label{Perturb}

\

\noindent
For $\hat{H}_{b}=-\vec{\si}\cdot\vec{\tilde{b}}(1+gz)+\frac{\bar{b}}{m_I}\si_z\hat{p}_z$,
it is difficult to find analytical solutions.
However, as $\hat{H}_{b}$ must be very tiny in the concordant frame \cite{Causality}, it can be regarded as perturbation.
Without this tiny perturbation, the energy levels must be degenerate for $\hat{H}_0$,
so the tiny perturbation only slightly split energy levels and it is proper to utilize the degenerate perturbation theory.
As already mentioned, we can ignore the slow sidereal time dependence for current experiments unless ultra-stable
polarized neutron beams is attainable.

For convenience, we separate $\hat{H}_{b}$ into $\hat{H}_{b}=\hat{H}_1+\hat{H}_2$, where $\hat{H}_1\equiv-\vec{\si}\cdot\vec{\tilde{b}}(1+gz)$
and $\hat{H}_2\equiv-\frac{\bar{b}}{m_I}\si_z\hat{p}_z$.
In the subspace spanned by $|n,\uparrow\rangle$ and $|n,\downarrow\rangle$ with
\bea\label{State0}
\langle{z}|n,\uparrow\rangle=\left(
                        \begin{array}{c}
                          1 \\
                          0 \\
                        \end{array}
                      \right)\phi_n(z),\quad
\langle{z}|n,\downarrow\rangle=\left(
                        \begin{array}{c}
                          0 \\
                          1 \\
                        \end{array}
                      \right)\phi_n(z),
\eea
where $\phi_n(z)$ is given by Eq.~(\ref{BounceEf}). The matrix elements are given by
\bea\label{MatrixEle}&&
\hspace{-21mm}\left(
  \begin{array}{cc}
    \langle{n,\uparrow}|\hat{H}_1|n,\uparrow\rangle & \langle{n,\uparrow}|\hat{H}_1|n,\downarrow\rangle \\
    \langle{n,\downarrow}|\hat{H}_1|n,\uparrow\rangle & \langle{n,\downarrow}|\hat{H}_1|n,\downarrow\rangle \\
  \end{array}
\right)=-|\vec{\tilde{b}}|\left[1+\frac{2}{3}gL_cx_{n+1}\right]
\left(
  \begin{array}{cc}
    \cos\th & \sin\th e^{-i\phi} \\
    \sin\th e^{i\phi} & -\cos\th \\
  \end{array}
\right),
\eea
where we have used the following integrals
{\small
\bea\label{I1-3}&&
\hspace{-24mm}
\int_0^{+\infty}dz\phi_n(z)\frac{d\phi_n(z)}{dz}=0,~
\int_0^{+\infty}dz\phi_n(z)z\phi_n(z)=\frac{2}{3}L_cx_{n+1},~
\int_0^{+\infty}dz\phi_n(z)z\frac{d\phi_n(z)}{dz}=-\frac{1}{2}.\nonumber
\eea
}
The matrix (\ref{MatrixEle}) can be diagonalized with eigen-vectors
$|+\rangle=\left(
              \begin{array}{c}
                \cos(\th/2)e^{-i\phi} \\
                \sin(\th/2) \\
              \end{array}
            \right)$ and
$|-\rangle=\left(
              \begin{array}{c}
                \sin(\th/2) \\
                -\cos(\th/2)e^{i\phi} \\
              \end{array}
            \right)$.
The corresponding eigen-energy to the leading order of $|\vec{\tilde{b}}|$ is
\bea\label{EigenEM}&&
\hspace{-15mm}\left(
  \begin{array}{c}
    E_{n~+} \\
    E_{n~-} \\
  \end{array}
\right)=
E_{n0}\pm\de{E}_n=\left[\frac{(m_{_G}g\hbar)^2}{2m_I}\right]^{1/3}x_{n+1}
\mp|\vec{\tilde{b}}|\left[1+\frac{2}{3c^2}gL_cx_{n+1}\right].
\eea
where we have inserted the $c^2$ and $\hbar$ for later calculations.
We note that $E_{n~+},~E_{n~-}$ in Eq.~(\ref{EigenEM}) are exactly the first-order approximation
of the eigen-energy $E_{n\pm}\mp B_0$ in Eq.~(\ref{ExactEn}),
since the extra term $-\bar{b}~\si_z\hat{p}_z$ does not contribute.
However, it does contribute to higher-order approximation of eigen-energies,
as $\int_0^{+\infty}dz\phi_n(z)\frac{d\phi_m(z)}{dz}\neq0$ for $n\neq{}m$.
Note that Eq.~(\ref{EigenEM}) leads to tiny corrections to the transition frequency between different gravitational bound states,
\bea\label{frequenpi}&&
\nu_{mn}^\pi=(E_{m~\pm}-E_{n~\pm})/h
=\nu_{0mn}\mp\frac{2g}{3hc^2}|\vec{\tilde{b}}|L_c(x_{m+1}-x_{n+1}),\\\label{frequensi}&&
\nu_{mn}^{\si\pm}=(E_{m~\mp}-E_{n~\pm})/h
=\nu_{0mn}\pm2\frac{|\vec{\tilde{b}}|}{h}\left[1+\frac{g}{3c^2}L_c(x_{m+1}+x_{n+1})\right],
\eea
where $\nu_{0mn}\equiv\left[\frac{(m_{_G}g\hbar)^2}{2m_I}\right]^{1/3}\frac{x_{m+1}-x_{n+1}}{h}$ is the conventional transition frequency without extra contributions,
and we have borrowed the notations $\pi$ and $\si\pm$ from atom optics to mark the spin-conserving and spin-flip transitions, respectively.
Here to make the results more transparent, we restored the $\hbar\,,c$ in this subsection.
Substituting the experimental results $\nu_{02}=(464.8\pm1.3)$Hz, $\nu_{03}=(649.8\pm1.8)$Hz
and the derived local acceleration $g=(9.866\pm0.042)\mathrm{m}/\mathrm{s}^2$ \cite{GRS} into Eq.~(\ref{frequenpi}), we get an upper bound
$|\vec{\tilde{b}}|<3.9\times10^{-3}$GeV, where we have already averaged the bounds obtained from the uncertainty of $\nu_{02},~\nu_{03}$.
Estimation from the energy resolution $\Delta{E}=2\times10^{-15}$eV \cite{GRS} gives $|\vec{\tilde{b}}|<1.3\times10^{-3}$GeV.
If the energy resolution is able to reach $\Delta{E}\sim10^{-21}$eV \cite{IWAN} in the future,
the bound can be improved by at least $6$ orders of magnitude, \ie, $|\vec{\tilde{b}}|<6\times10^{-10}$GeV.
If polarized neutron beam is attainable \cite{IWAN} and spin-flip transition frequency can be measured with the same precision around $1.3\sim1.8$Hz in the near future, a bound $|\vec{\tilde{b}}|<3.2\times10^{-24}$GeV can be obtained
from the $\si\pm$-transition in Eq.~(\ref{frequensi}).
We also note that the rough estimate can be further refined by taking experimental details into account and
properly dealing with all known systematic errors. However, this is beyond the scope of this work.

\section{Summary}\label{PersDis}
In this paper, we calculate the tiny CPT-violating (CPTV) effects on neutrons' gravitational bound states in the minimal SME \cite{SMEG}.
Following the spirit of Ref.~\cite{FermiNonmini}, we redefine the LV coefficients and rewrite the spin-dependent LV Hamiltonian
(\ref{SpinLV}) in the uniform gravitational field \cite{YuriEPI}.
The main LV operators we concerned are in the Hamiltonian (\ref{VerticalHLV}).
The $\tilde{b}$ coefficient can be regarded as the neutron $\tilde{b}^n$ defined in Ref.~\cite{AKCL} with leading-order gravity corrections,
indicated by the hat in its definition, see Eqs.~(\ref{bLVeff1}--\ref{dLVeff}).
Though $\tilde{b}^n$ has already been tightly constrained by the comagnetometer experiment \cite{comagnetometer},
the $\tilde{b}$-term we considered represents LV free neutron spin-gravity couplings, and has only been weakly constrained.

First we obtain the operator equation of motion (\ref{HeisenEqn}) for spin precession from the Hamiltonian (\ref{HeffBLV}),
where the tiny LV correction to the Larmor frequency has been obtained.
The tiny (precession) frequency correction depends in a complicated way on both position and momentum,
and may be testable in the comagnetometer experiment \cite{comagnetometer}.
The $\hat{p}$-dependence of $\de\vec{\omega}_L$ may indicate that the tiny LV correction to the precession frequency increases with
increasing energy, and hence the LV spin-dependent effect may be enhanced in the high energy region,
and the LV spin-gravity effects in a relativistic situation could be an interesting topic to study in the future.

As we concerned most is the gravitational resonance spectroscopy experiment \cite{GRS}, we average out the
horizontal motion to get Eq.~(\ref{VerticalHLV}) with the ansatz in Ref.~\cite{Escobar1}.
With the parameters chosen from Ref.~\cite{GRS}, we give a rough estimate of various operators in Eqs.~(\ref{VerticalHLI}) and (\ref{VerticalHLV}), for details, see Table~\ref{EEO}. Finally we focus on the LV operators in Eq.~(\ref{EffLVH}).
To obtain analytical solutions as case studies, we consider $\tilde{b}$- and $\bar{b}$-terms separately.
For the $\tilde{b}$-term, we show that the precession measurement of the asymmetry in finding particles in perpendicular directions to the initial spin polarization may be a potential way in searching for LV.
The asymmetry (\ref{asymSid}) is proportional to $|\tilde{b}|$ at leading order,
and apart from a sidereal time dependence, it also increases with time in several days for ultra-stable polarized neutrons.
The time-dependent feature of the asymmetry is shown in Fig.~\ref{Asymmetry} and Fig.~\ref{ProbX}.
For clarity, we also show the spin evolution on the Bloch sphere in Fig.\ref{SpinX}.
For the $\bar{b}$-term, the inseparability of the spin-momentum coupling shows an external motion dependence for
the spin expectation values; see Eq.~(\ref{SpinVEV}). Further detailed study for more complicated analytical models
is also possible, but will be left for the future.

Finally, we calculate the CPTV coupling induced frequency shifts (\ref{frequenpi}) and (\ref{frequensi}) with
perturbation theory.
From the precisely measured spin-insensitive transition frequency between different gravitational bound states \cite{GRS},
we obtain a rough bound $|\vec{\tilde{b}}|<3.9\times10^{-3}$GeV.
Though this bound is much weaker than the neutron bounds of $\tilde{b}^n$ in comagnetometer and
comaser experiments (see Table D12 in Ref.~\cite{DATA}; Neutron sector), this is one of the few bounds on LV coefficients for
free neutron spin-gravity couplings \cite{IWAN}.
If the spin-flip transition frequency can be measured with comparable accuracy for polarized neutron beam in future,
the bounds can be significantly improved.

Aside from the GRS experiments, the CPTV spin-gravity couplings can also lead to violation of the universal free fall (UFF),
indicated in the main context.
Actually, if we naively interpret the potential of spin-$\hf$ $^{87}\mathrm{Sr}$
as $V=(m_{\mathrm{Sr}}-\vec{\si}\cdot\vec{\tilde{b}})(1+\Phi)$ and assume no exotic corrections to spin-0 $^{88}\mathrm{Sr}$,
from the E$\ddot{\mathrm{o}}$tv$\ddot{\mathrm{o}}$s parameter $\eta=(0.2\pm1.6)\times10^{-7}$ \cite{Tino2014}, we can get $|\vec{\tilde{b}}|<4\times10^{-8}m_{\mathrm{Sr}}\simeq3.3\times10^{-6}$GeV, consistent with the bound $|\vec{\tilde{b}}|<2.5\times10^{-8}m_{\mathrm{Sr}}$ extracted from the spin-gravity coupling strength $k=(0.5\pm1.1)\times10^{-7}$.
Note that unlike the bound in Sec.~\ref{Perturb}, this more stringent bound is for the nuclear bound state of neutron, though it is still
much weaker than the corresponding bound reported in Ref.~\cite{DATA}.
Anyway, it still indicates that more stringent test of equivalence principle with polarized matter in future \cite{HUSTS} may
be able to provide much tighter bounds on LV spin-dependent coefficients, such as $\tilde{b}$.

Actually, besides gravitational bound states, the scattering states in the tunneling region \cite{ZXBterm}\cite{ZXTT}
are also interesting for study.
For the LV spin-dependent effects, this work only provides a simple study of $b$-type coefficients
with neutrons' gravitational bound states.
A large set of LV spin-gravity couplings are still open to exploration.

\section{Acknowledgments}
We would like to appreciate the valuable encouragement and helpful discussions with M. Snow,
A. Kosteleck\'y and D.F. Zeng, and also to express the gratitude to D. Colladay and J. Long in discussion.
ZX was partially supported by National Science Foundation of China (11605056, 11875127, 11974108)
and the Fundamental Research Funds for the Central Universities under No. 2017MS052.
LS was supported by the National Natural Science Foundation of China
(11975027, 11991053), and the Young Elite Scientists Sponsorship Program
by the China Association for Science and Technology (2018QNRC001).

\appendix
\section{Various operators in tracing over the horizontal motion}\label{VerticalO}
The effective vertical operator is obtained as
$\langle\hat{O}\rangle\equiv\int{d\vec{r}_\perp}\psi^*(\vec{r}_\perp)\hat{O}\psi(\vec{r}_\perp)$,
where $\psi(\vec{r}_\perp)\equiv\frac{1}{\sqrt{\pi}\si}\exp \left({i\vec{p}_\perp\cdot\vec{r}_\perp-\frac{\vec{r}_\perp^2}{2\si^2}}\right)$ represents the horizontal wave-packet \cite{Escobar1}.
The various terms after tracing over the horizontal d.o.f. are
\bea&&\label{PhiAver}
\hspace{-15mm}\langle\Phi\rangle=\langle(1+gz)\rangle=1+gz,\quad
\left\langle\frac{\hat{\vec{p}}^2}{2m}\right\rangle=\frac{\hat{p}_z^2}{2m}+\left[\sum_{a=1}^2\frac{p_a^2}{2m}+\frac{1}{\si^2}\right],\nonumber
\eea
\bea&&
\hspace{-15mm}\left\langle\Phi\frac{\hat{\vec{p}}^2}{2m}\right\rangle=(1+gz)\left\langle\frac{\hat{\vec{p}}^2}{2m}\right\rangle,\quad
\left\langle-\frac{i}{2m}\vec{\nabla}\Phi\cdot\hat{\vec{p}}\right\rangle=-\frac{i}{2m}\vec{g}\cdot\langle\hat{\vec{p}}\rangle
=-\frac{i}{2m}g\hat{p}_z,\nonumber
\eea
\bea&&
\hspace{-15mm}\left\langle\frac{1}{2m}\vec{\si}\cdot(\vec{\nabla}\Phi\times\hat{\vec{p}})\right\rangle=-\frac{g}{2m}\cdot(\vec{\si}\times\vec{p})_z,\quad
\left\langle\frac{\Phi}{2m}\vec{\si}\cdot(\vec{\nabla}\Phi\times\hat{\vec{p}})\right\rangle=-\Phi\frac{g}{2m}\cdot(\vec{\si}\times\vec{p})_z,
\nonumber
\eea
\bea&&
\hspace{-15mm}\langle\vec{\si}\cdot\vec{\tilde{b}}(1+\Phi)\rangle=(1+gz)\vec{\si}\cdot\vec{\tilde{b}},\quad
\langle(1+\Phi)\vec{\si}\cdot\hat{\vec{p}}\rangle=(1+gz)\left[\si_z\hat{p}_z+\sum_{a=1}^2\si_ip_i\right],\nonumber
\eea
where we have chosen the local coordinates such that the $z$ axis is along the vertical direction,
which is anti-parallel to the direction of the local gravitational acceleration $\vec{g}$.

\end{document}